# Networked Multiagent Safe Reinforcement Learning for Low-carbon Demand Management in Distribution Network

Jichen Zhang, Linwei Sang, Yinliang Xu, *Senior Member IEEE*, and Hongbin Sun, *Fellow, IEEE*

*Abstract*—This paper proposes a multiagent based bi-level operation framework for the low-carbon demand management in distribution networks considering the carbon emission allowance on the demand side. In the upper level, the aggregate load agents optimize the control signals for various types of loads to maximize the profits; in the lower level, the distribution network operator makes optimal dispatching decisions to minimize the operational costs and calculates the distribution locational marginal price and carbon intensity. The distributed flexible load agent has only incomplete information of the distribution network and cooperates with other agents using networked communication. Finally, the problem is formulated into a networked multi-agent constrained Markov decision process, which is solved using a safe reinforcement learning algorithm called consensus multi-agent constrained policy optimization considering the carbon emission allowance for each agent. Case studies with the IEEE 33-bus and 123-bus distribution network systems demonstrate the effectiveness of the proposed approach, in terms of satisfying the carbon emission constraint on demand side, ensuring the safe operation of the distribution network and preserving privacy of both sides.

## I. INTRODUCTION

DEMAND side management has been widely applied to deal with the increasing integration of renewable energy resources and growing load consumption [1]. Incentive-based and price-based demand side response are effective ways to encourage consumers to participate in the power market including energy market and ancillary service market [2].

Trojani *et al.* show an example of security-constrained unit commitment using DERs and flexible loads [3]. Wei *et al.* propose a bi-level mathematical model for the industrial users based on distribution locational marginal price (DLMP). which is solved through single-level reformulation [4]. Tindemans *et al.* introduce a decentralized thermostatically control method for aggregate load for the flexible demand management [5]. A localized demand management is proposed in [6] to cope with the challenges of peak load, overloads and voltage violations. In [7], Lyapunov optimization is applied for flexible loads in the real-time operation of distribution network. Zhang *et al.* develop the modelling and algorithms for locally communicated and time-varying dynamic distribution system. Therefore, distributed flexible demand management has been considered in the operation of distribution network and case studies validate their effectiveness in the distribution network operation [8], [9]. However, most researches neglect the privacy preservation between the load user and system operator, since they usually belong to different stakeholders and complete information cannot be shared. The cooperative operation framework considering information privacy, which can facilitate the active participation of flexible loads in the distribution network, still needs to be investigated.

Recently, carbon emission reduction has also raised great concerns in power systems [10]. Researches have focused on the low-carbon operation methods for the power systems such as carbon taxes and allowance on the generation side. Yan *et al.* introduce an optimized-stepped carbon trading mechanism in the seasonal integrated energy system low-carbon economic dispatch framework to reduce the operational costs and carbon emission [11]. However, consumers should also be responsible for the carbon emission. Carbon emission flow introduced in [12] is used to allocate the carbon emission responsibility to the demand side. Wang *et al.* propose a two-stage scheduling demand side management framework in both electricity and carbon market with double carbon tax mechanism [13]. Cheng *et al.* study energy-carbon integrated prices in the coordination of transmission-level and distribution-level to realize the low-carbon operation in multi energy systems [14]. Huo *et al.* introduce a spatio-temporal flexibility requirement envelope for the low-carbon power system operation with uncertainties [15]. Yan *et al.* consider a bi-level carbon trading scheme to realize the maximization of consumer surplus and incentive the low-carbon operation [16]. Despite that low-carbon operation has been studied both on generation side and demand side through the carbon emission flow, researchers mainly focus on the specific low-carbon operation model and have developed some methods to address the nonlinearities. However, due to lacking of detailed and complete information of the carbon emission flow, the demand side low-carbon management with the carbon intensity specified on different nodes cannot be handled properly. Demand management under unknown model with coupled power and carbon flow needs to be further investigated.

Reinforcement learning (RL), known as a novel data-based algorithm, has been widely used in the power system researches along with the advancement of the data and communication technologies. Increasing number of distributed energy resources not only incur the high computational burden, but also pose great treat to the distribution network operation due the uncertainty. RL is closely related to the optimal control and dynamic programming, showing great edges on the optimization of systems with inaccurate or even no models, which is known as model-free algorithms. It has played an important role in the energy management field [18]-[20]. For distributed settings in distribution system, multi-agent reinforcement learning has been forwarded to control the frequency and voltage in the real-time operation [17], [21], [28]. However, safe operation calls for great concerns when we applying reinforcement learning in power systems. Achiam *et al.* propose an algorithm called constrained policy optimization which can find the optimal policy update in the safe limits [22]. They have been applied in the distribution network or microgrids to deal with the large-scale integrated renewable energy resources and voltage violations [23], [24]. However, algorithms always use a single value function or share the same state value function in multiagent RL which are basically trained in a centralized way [25]. They may suffer from heavy communication burdens, which is prone to communication failures. Moreover, few researches focus on the adaptivity of strategies to the time-varying and uncertain renewable energy resources and loads for the low-carbon demand management in the distribution network.

To tackle with the above-mentioned challenges, this paper proposes a multiagent based bi-level operation framework for the low-carbon demand management in the distribution network considering the privacy preservation between load agents and the distribution network operator, which is also adaptive to the uncertain renewable energy resources. In the upper level, the load agents, which are subject to the local carbon emission allowance, make control decision to maximize their profits based on the electricity price signal; in the lower level, the distribution system operator aims to minimize the overall system operation cost. The distribution network operator derives the distribution locational marginal price and nodal carbon intensity through the optimal power flow calculation, which are broadcast to the load agents. To obtain the optimal adaptive strategies for the profit maximization of all cooperative load agents, consensus multi-agent constrained policy optimization is proposed, which is proven to be effective and computational efficient.

The main contributions of this paper are summarized below:

(1) Compared with the demand side management framework in [3]-[9] which ignore the different interests of loads and distribution network operator and carbon reduction, a bi-level operation framework for networked load agents is proposed to maximize the profit of aggregate loads within the carbon allowance and minimize the cost of distribution network considering privacy preservation.

(2) Contrary to [4], [14], [16] that dealing with the nonconvexities of the bi-level operation framework through single level reformulation or decomposition, the developed distributed demand management model is reformulated into a networked multiagent constrained Markov Decision Process which only needs partial information exchange, decouples intractable power-carbon flow constraints and adapts to variant states.

(3) Different from the safe reinforcement learning algorithms developed in [19]-[23] which need a centralized controller in the training process, this paper proposes a novel consensus multiagent constrained policy optimization approach to realize the optimal coordination between flexible loads and the distribution network, while satisfying the carbon emission limit and preserving the private information.

This paper is organized as follows: Section II introduces the problem formulation for the operation framework. In Section III, the proposed consensus multi-agent constrained policy optimization is presented. Simulation studies are conducted in Section IV. Finally, conclusions are drawn in Section V.

## II. PROBLEM FORMULATION

The section introduces the problem formulation for the optimal bi-level operation of the distribution network with flexible loads considering the demand-side carbon emission allowance. Firstly, we introduce the bi-level distribution network operation framework and the CEF calculation. Then, we convert the formulated model into a networked multi-agent constrained Markov decision process (NMACMDP). The overall operation framework and the detailed information exchange is illustrated in Fig. 1.

### A. Distribution Network Operation Framework

Flexible loads are important demand-side resources for the active operation of the distribution network. They can be divided into different types according to the characteristics including uncontrollable loads, adjustable loads and transferable loads. In the proposed framework, various types of flexible loads are aggregated, which belong to a single stakeholder. The aggregate load agents make decision in a distributed way to maximize their profits including the utility and payments to distribution system operator in the upper level. Then the system operator minimizes the total cost by making the dispatching plans of the distributed generators in the lower level. The load agents send the demand and receive the price signals and carbon intensity from distribution system operator. Note that the privacy of cost or utility needs to be preserved between the aggregated loads and system operator so that the optimized information cannot be exchanged due to their different interests. Through the demand side management, we could reduce the operation cost and improve the utility of load consumption when confining the carbon emission in the demand side. The specific bi-level mathematical model for the loads and distribution operator is given below.

(1) *Upper Level: Aggregated Loads:* The aggregated loads include three types whose detailed model is shown below:

*Uncontrollable loads:* The uncontrollable loads are electrical devices that are essential or urgent such as lights in the households, computational and storage servers in the companies or surgical equipment in hospital. These loads combined together are regarded as non-flexible which cannot be adjusted. In the problem, we assume that uncontrollable loads, denoted by $P^i_{load,uc,t}$ are sampled from real-time data at the beginning of each time step.

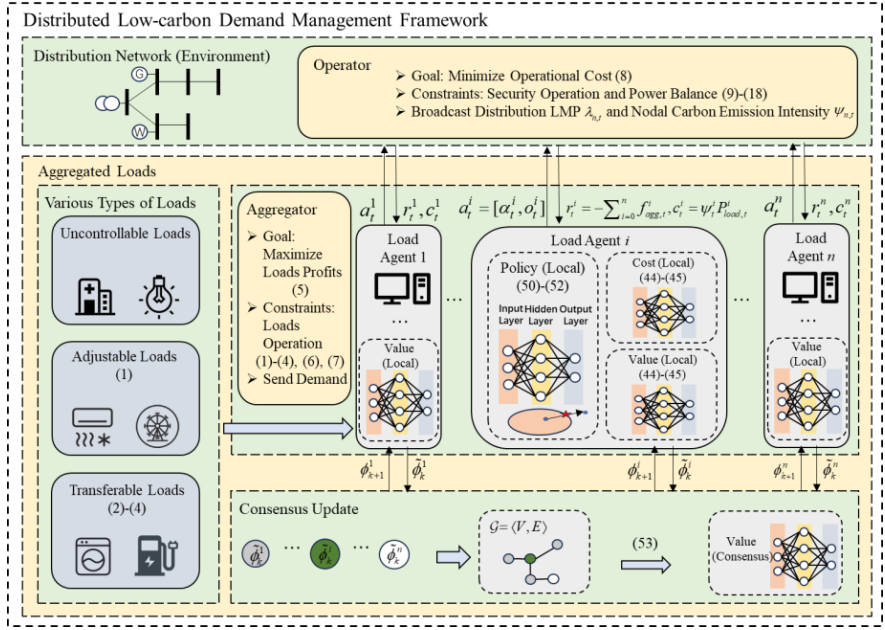

Fig. 1 Distributed low-carbon demand management framework

*Adjustable loads:* The controllable loads are electrical devices that can be manipulated to fulfill consumers' desire such as air conditions and televisions in the households or other entertainment infrastructure. The loads can be adjusted by a load controller in a specific range which is fixed at a certain bus in the distribution network. The increasing of load consumption contributes to the improvement of consumers' satisfaction so that the utility functions are introduced to maximize the social welfare from demand side. The mathematical model of controllable loads can be formulated as below:

$$P^i_{load,a,t} = \alpha^i_t P^i_{load,a,t,\max}, \alpha^i_t \in [0,1] \quad (1)$$

where $\alpha^i_t$ is controllable parameter for adjustable load for agent $i$ at time step $t$.

*Transferable loads:* The transferable loads are electrical devices whose total operational time is fixed and the power consumption for a specific timestep is a constant such as washing machine in the household. They can be transferred from one time interval to another based on the load agent controller so that the state can be set as 'on' or 'off' at a specific timestep. Notably, the loads must be arranged to operate for certain times otherwise they will be punished. The mathematical model of controllable loads can be formulated as:

$$P^i_{load,tr,t} = o^i_t P^i_{load,tr,t}, o^i_t \in \{0,1\} \quad (2)$$

$$\prod_{t_s}^{t_s+\Delta_i} o^i_t = 1, t_s \in \{1,2,\ldots,T\} \quad (3)$$

$$\sum_0^T o^i_t = \Delta_i \quad (4)$$

where $o^i_t$ is controllable parameter for transferable load for agent $i$ at time step $t$, $t_s$ is start time for transferable load and $\Delta_i$ is the last running time for transferable load for agent $i$. For each nodal controller, it decides the actions for all the different types of loads and submit the local total demand to the operator. Then the nodal price is cleared according to the next part. Finally, the aggregated loads can maximize their summations of welfare with the following optimization problem including the utility function of adjustable loads and minimize the expenses:

$$f_{agg} = \sum_t^T \sum_i^I (a^i_{load}(P^i_{load,a,t})^2 + b^i_{load} P^i_{load,a,t}) - \lambda_{n_i,t} P_{load,n,t} \quad (5)$$

s.t. $P_{load,n_i,t} = P_{load,uc,i,t} + P_{load,a,i,t} + P_{load,tr,i,t} \quad (6)$

$$\sum_t^T \psi_{n_i,t} P_{load,n_i,t} \leq EM_{quota,i} \quad (7)$$

(1)-(4)

where $n_i$ is the bus with a local flexible load agent.

(2) *Lower Level: Distribution System Operator:* In the proposed framework, the goal of the operator in the lower level is to minimize the operational cost for the distributed generator including diesel generators and gas power plant and renewable energy resources including wind turbines and photovoltaic and formulate the operational cost functions and constraints into an optimization problem. Firstly, the detailed objective of the distribution network operation is shown below:

$$\min f_{DNO} = \sum_t^T \sum_n^N a_{DG,n} P^2_{DG,n,t} + b_{DG,n} P_{DG,n,t} + c_{DG,i} + k_{re,n} P_{re,n,t} + \pi_{n,t} P_{grid,n,t} \quad (8)$$

The first three terms represent the quadratic, linear and constant coefficient of the cost for the distributed generators respectively. The fourth term indicates that the cost for the renewable generators and the last term represents the fees paid to the external grid by the distribution network operator.

Also, the security of operation needs to be guaranteed so that the following constraints should be considered in the model:

$$P_{DG,n,\min} \leq P_{DG,n,t} \leq P_{DG,n,\max} \quad (9)$$

$$0 \leq P_{re,n,t} \leq P_{re,n,t,\max} \quad (10)$$

$$0 \leq P_{grid,n,t} \leq P_{grid,n,t,\max} \quad (11)$$

$$P_{n,t} = P_{grid,n,t} + P_{DG,n,t} + P_{re,n,t} - P_{load,n,t} : \lambda_{n,t} \quad (12)$$

$$P_{nm,t} = P_{n,t} + \sum_{l:(m,l) \in L} P_{ml,t} + r_{nm} I_{nm,t} \quad (13)$$

$$Q_{nm,t} = Q_{n,t} + \sum_{l:(m,l) \in L} Q_{ml,t} + x_{nm} I_{nm,t} \quad (14)$$

$$v_{m,t} = v_{n,t} - 2(r_{nm} P_{nm,t} + x_{nm} Q_{nm,t}) + (r_{nm}^2 + x_{nm}^2) I_{nm,t} \quad (15)$$

$$\left\| \begin{matrix} 2P_{nm,t} \\ 2Q_{nm,t} \\ I_{nm,t} - v_{n,t} \end{matrix} \right\| \leq I_{nm,t} + v_{m,t} \quad (16)$$

$$\underline{v} \leq v_{n,t} \leq \overline{v} \quad (17)$$

$$I_{mn,t} \leq \overline{I} \quad (18)$$

where $m$, $n$ is the index for buses in the distribution network. Eqn(9) represent the magnitude constraint of distributed generators. Eqn(10) shows that power output of renewable energy resources cannot exceed the limit of real-time maximum power. Eqn(12) denotes the nodal power balance constraints and the multipliers for the constraints denotes the distribution locational marginal price $\lambda_{n,t}$ for the nodal loads. Then the price signals are broadcast to the aggregated loads. Eqn(13)-(18) represents the Distflow model constraints for the distribution network after second order cone relaxation from branch flow model where $v_{n,t}$ and $I_{mn,t}$ are the square of nodal voltage and line current from bus $m$ to bus $n$ at time step $t$. Note that the load is submitted to the operator in the first stage so that the optimization problem known as second order cone programming can be solved by commercial solvers such as Gurobi.

(5) *Carbon Emission Flow:* To transfer the carbon emission responsibility to the demand side, we need to calculate the nodal carbon intensity for flexible loads in the distribution network so that the CEF calculation is introduced based on the optimal power flow.

The nodal carbon intensity in the distribution network can be defined as the following expression:

$$P_{in,n,t} = P_{grid,n,t} + P_{DG,n,t} + P_{g,n,t} + P_{re,n,t} \quad (19)$$

$$EM_{in,n,t} = P_{grid,n,t} \psi_{grid,t} + P_{DG,n,t} \psi_{DG,n,t} + P_{g,n,t} \psi_{g,n,t} + P_{re,n,t} \psi_{re,n,t} \quad (20)$$

$$\psi_{n,t} = \frac{\sum_{l:(l,n) \in L^+} P_{ln,t} \psi_{l,t} + EM_{in,n,t}}{\sum_{l:(l,n) \in L^+} P_{ln,t} + P_{in,n,t}} \quad (21)$$

Eqn(19) represents the sum of nodal power injection $P_{in,n,t}$ and Eqn(20) denotes the sum of nodal carbon emission $EM_{in,n,t}$ corresponding with the power injection. Then we can calculate the nodal carbon intensity by Eqn(21). Note that the carbon intensity for distributed generators is pre-defined while branch intensity relevant with inflow active power is required to be calculated iteratively. Matrix forms can simplify the calculation. Firstly, we define the branch power flow matrix as $\boldsymbol{P}_t^B$, and the entry of the matrix is determined as follows

$$\boldsymbol{P}_t^B = \begin{cases} P_{nm,t}, & \text{if positive power flow from } n \text{ to } m \\ 0, & \text{otherwise} \end{cases} \quad (22)$$

Note that $\boldsymbol{P}_t^B$ is an N-level matrix and the diagonal elements are zero. In the carbon emission flow, the outflow power don't affect the nodal carbon intensity. Therefore, Eqn(19)-(21) can be transformed into the matrix form as:

$$\boldsymbol{EM}_t = (\boldsymbol{P}_t^B)^T \cdot \boldsymbol{\Psi}_t^N + \boldsymbol{P}_t^G \cdot \boldsymbol{\Psi}_t^G. \quad (23)$$

$\boldsymbol{\Psi}_t^N$ is an N-dimensional vector for nodal carbon intensity. $\boldsymbol{P}_t^G$ is an N × Z level matrix with the set $G$ of generators. If the $z$th distributed generator is connected to the $n$th bus, the element in the $n$th row and $z$th column is $P_t^G$; otherwise, it will be set to zero. $\boldsymbol{\Psi}_t^G$ is a Z dimensional column vector in which the $z$th element is carbon intensity of the $z$th generator. For the nodal power influx in the denominator of (21), we define a diagonal matrix $\boldsymbol{P}_t^N$ as follows:

$$\boldsymbol{P}_t^N = \mathrm{diag} \left\{ \zeta_{N+Z} \cdot \begin{bmatrix} \boldsymbol{P}_t^B \\ (\boldsymbol{P}_t^G)^T \end{bmatrix} \right\}. \quad (24)$$

where $\zeta_{N+Z}$ is a unit N+Z dimensional row vector. Then we can rewrite the (21) into the matrix form as follows:

$$\boldsymbol{P}_t^N \cdot \boldsymbol{\Psi}_t^N = (\boldsymbol{P}_t^B)^T \cdot \boldsymbol{\Psi}_t^N + \boldsymbol{P}_t^G \cdot \boldsymbol{\Psi}_t^G. \quad (25)$$

Then the nodal carbon intensity can be calculated as follows:

$$\boldsymbol{\Psi}_t^N = \left( \boldsymbol{P}_t^N - (\boldsymbol{P}_t^B)^T \right)^{-1} \cdot \left( \boldsymbol{P}_t^G \cdot \boldsymbol{\Psi}_t^G \right). \quad (26)$$

To sum up, in the operation framework, the aggregated distributed flexible loads aim to maximize their profits under the limits of carbon emission allowance in the upper level and the distribution system operator aims to minimize the total generation cost when faced with uncertain renewable energy resources and uncontrollable loads in the lower level. The privacy is protected between different interests so that the aggregated loads only transfer demand information and the distribution operation operator only broadcasts the distribution local marginal price and carbon intensity. Note that the loads with carbon emission allowance are operated in a distributed way and the coupled electricity and carbon operation with nonlinearities is intractable. Therefore, we introduce the following Markov decision process to formulate the problem and solve it.

### B. Networked Multi-Agent Constrained MDP

Based on the theory suggested in [27], we introduce the networked multi-agent constrained Markov decision process with a tuple $\langle \mathcal{N}, \mathcal{S}, \mathcal{A}, \mathrm{p}, \rho^0, \gamma, \boldsymbol{R}, \boldsymbol{C}, d, \mathcal{G} \rangle$. Here, $\mathcal{N} = \{1,...,I\}$ is the set of agents, $\mathcal{S}$ is the global state space, $\mathcal{A} = \prod_{i=0}^{I} \mathcal{A}^i$ is the product of the agents' action spaces $\mathcal{A}^i$ which is the joint action space of multi-agents, p is the probabilistic transition function, $\rho^0$ is the initial state distribution, $\gamma$ is the discount factor, $\boldsymbol{R} = \{r_t^i\}_{i=0}^I$ represents the set of n agents' local reward function $r_t^i$, $\boldsymbol{C} = \{c_t^i\}_{i=0}^I$ stands for the set of n agents' local cost function $c_t^i$, $d$ stands for the corresponding cost limits and finally a communication network $\mathcal{G} = \langle V, E \rangle$. In the NMACMDP, the agents in the set select their actions $a_t^i$ at time step $t$ based on the states $s_t$ they are in according to the policy $\pi^i(a^i | s)$ which is defined as the mapping from global states to local actions. Combining with other agents' actions, a joint action $a_t$ is given to the environment. Then each agent receives a local reward signal $r_t^i$ from the environment together with a local cost signal $c_t^i$. Meanwhile, the environment transits to a new state $s_{t+1}$ according to the transition probability $\mathrm{p}$. In the process, the agent can communicate with its neighbors define in the communication graph $\mathcal{G}$ to share partial local information. Note that the transition probability distribution which is unknown in the model-free reinforcement learning in the interaction process that satisfies Markov property, that is, the value or probability of occurrence of state and reward only depends on the previous state and corresponding action, not related to earlier states and actions.

In our environmental setting, we consider a cooperative setting that all agents share the same reward signal, defined as $R_t$, and aim to find the optimal joint local control policy $\pi(a|s) = \prod_{i=0}^{I} \pi^i(a^i|s)$, such that it maximizes the expected cumulative rewards

$$J(\pi) = \mathbb{E}_{s\sim \mathrm{p}, a\sim \pi}[\sum_{t=0}^{T} \gamma^t R(s_t, a_t)] \quad (27)$$

satisfying each agent's safety constraints, denoted as

$$J_C^i(\pi) = \mathbb{E}_{s\sim \mathrm{p}, a\sim \pi}[\sum_{t=0}^{T} \gamma^t C^i(s_t, a_t^i)] \leq d_t^i \quad (28)$$

When the constraints are satisfied, the joint policies are referred to as feasible. There are two important functions in the process, global state value function and global action value function showing below:

$$Q_\pi(s,a) = \mathbb{E}_{s\sim \mathrm{p}, a\sim \pi}[\sum_{t}^{\infty} \gamma^t R(s_t, a_t)|s_t = s, a_t = a] \quad (29)$$

$$V_\pi(s) = \mathbb{E}_{a\sim \pi}[Q_\pi(s,a)] \quad (30)$$

which is defined as the estimation of the expected, accumulative, discounted reward for the state $s_t$ and for the action $a_t$ at timestep t. Then for the multi-agent state-action value function. Similarly, the local state cost value function and local action cost value function are defined for expected, accumulative, discounted cost as below:

$$Q_{C,\pi}^i(s,a^i) = \mathbb{E}_{s\sim \mathrm{p}, a\sim \pi}[\sum_{t}^{\infty} \gamma^t c^i(s_t, a_t^i)|s_t = s, a_t^i = a^i] \quad (31)$$

$$V_{C,\pi}^i(s) = \mathbb{E}_{a\sim \pi}[\sum_{t}^{\infty} \gamma^t c^i(s_t, a_t^i)|s_t = s] \quad (32)$$

On top of them, we introduce the advantage function for the reward and cost which is essential in the proposed algorithm.

$$A_\pi(s,a) = Q_\pi(s,a) - V_\pi(s) \quad (33)$$

$$A_{C,\pi}^i(s,a_i) = Q_{C,\pi}^i(s,a_i) - V_{C,\pi}^i(s) \quad (34)$$

*C. Formulate Flexible Demand Management as an NMACMDP*

In the proposed operation framework, the distributed load controllers equipped at each bus in the distribution network are regarded as multi-agents in the reinforcement learning. Based on the perquisites, we formulate the flexible load control problems in the active management of distribution network into an NMACMDP.

(1) *State:* The state of the environment is defined as $s_t = [t, \boldsymbol{P}_{load,uc,t}, \boldsymbol{P}_{re,t,\max}, \boldsymbol{o}_{t-1}, \Delta_{l,t}, \boldsymbol{EM}_{t-1}]$, and $t$ is the current time in one dispatch circle, $\boldsymbol{P}_{load,uc,t}$ is the vector of real-time maximum power output of renewable energy generators equipped in the distribution network, $\boldsymbol{P}_{re,t,\max}$ is the vector of actual uncontrollable load consumption at time step $t$, $\boldsymbol{o}_{t-1}$ is the state of transferable loads at time step $t$-1, $\Delta_{l,t}$ represents the time of duration when the transferable loads have run and $\boldsymbol{EM}_{t-1}$ is the cumulative carbon emission of the aggregated loads at time step $t$..

(2) *Action:* The actions of each local controller are defined as $a_t^i = [\alpha_t^i, o_t^i]$. $\alpha_t^i$ is the continuous decision variable for the adjustable loads and $o_t^i$ is the discrete decision variables for the transferable loads. Note that the $o_t^i$ has two states, including 'on' and 'off' which means that the loads can be set on or off at each time step when satisfying their operational constraints. The joint actions for the environment are defined as $\boldsymbol{a}_t = [a_t^0, a_t^1, \ldots, a_t^n]$.

(3) *Environment:* The distribution system operator is set as the environment for the NMACMDP. It receives the load consumption from the aggregate load and regard it as a constant in the optimization problem (8)-(18). Then the operator in the lower level can minimize the operation cost of the system. The problem is convex so that we can solve it by commercial solvers and derive the distribution locational marginal price and carbon intensity.

(4) *Reward:* The reward signals are set as the objective of the welfare of aggregated loads maximization optimization problem at time step $t$. Due to the cooperative settings for load control agents, the rewards for them are the same, defined as:

$$r_t^i = -\sum_{i=0}^{n} f_{agg,t}^i \quad (35)$$

(5) *Cost:* The cost signals for each load controller agent are defined as the carbon emission of the nodal loads and the limitations for the costs are the carbon emission allowance set before. The agents can obtain the cost signals through the total load consumption and the nodal carbon intensity calculated from the power flow. Note that the cost signals and cost limits are local and private for each controller, defined as:

$$c_t^i = \begin{cases} \sum_{t=0}^{T} \psi_t^i P_{load,t}^i, & t = T \\ 0, & \text{else} \end{cases}, \quad d_t^i = \begin{cases} EM_{quota}^i, & t = T \\ 0, & \text{else} \end{cases} \quad (36)$$

Through the reformulation, the demand side carbon intensity calculation can be decoupled with power flow in the optimization. Then the NMACMDP can be solved by the proposed algorithm in the next section.

### III. METHODOLOGY

In the section, we will introduce the proposed on-policy deep reinforcement learning algorithm for the operation problem. Firstly, we will give a preliminary on constrained policy optimization. Then the consensus multi-agent constrained policy optimization is introduced to solve the proposed problem.

*A. Constrained Policy Optimization*

In the real physical system, the state space and action space are often infinite, which will suffer severely from the curse of dimensionality. Deep neural network (DNN) can work as a function approximation to represent the policy which is known as actor network, parameterized by $\theta$ and state value function which is known as critic network, parameterized by $\phi$. The local policy search can be applied in the train of the actor network to search for a feasible and optimal policy. In the constrained MDPs, the iterative update optimization for maximizing $J(\pi)$ within a local neighborhood of the past iteration can be illustrated as:

$$\pi_{k+1} = \arg\max J(\pi) = \arg\max \mathbb{E}[\sum_{t=0}^{T} \gamma^t R(s_t, a_t)]$$
$$\text{s.t. } J_C(\pi) = \mathbb{E}[\sum_{t=0}^{T} \gamma^t C(s_t, a_t)] \leq d \quad (37)$$
$$D(\pi, \pi_k) \leq \delta$$

where $D$ represents the distance between the distributions, $\delta$ is the step size for the update constraint and $d$ is the cost limits. The optimization problem is difficult to practice so that we derive the trust region policy optimization (TRPO) to approximate the update. In the approximation, the objective and constraints are replaced by the surrogate functions which is estimated from the samples collected from the current policy. The KL-divergence is applied to replace the distance of distributions. Through the trust region policy iteration, the lower bound performance improvement can be guaranteed based on the following expression:

$$J(\pi_k) - J(\pi) \geq$$
$$\frac{1}{1-\gamma} \mathbb{E}_{s\sim p, a\sim \pi_k} \left[ \begin{array}{c} A_\pi(s,a) - \\ \frac{2\gamma \max[\mathbb{E}_{a\sim \pi_k}[A_\pi(s,a)]}{1-\gamma} \sqrt{\frac{1}{2} D_{KL}(\pi_k \| \pi)[s]} \end{array} \right] \quad (38)$$

Similarly, we have equivalent expressions for the costs which can ensure the safety of the policy improvement as:

$$J_C(\pi_k) - J_C(\pi) \leq$$
$$\frac{1}{1-\gamma} \mathbb{E}_{s\sim p, a\sim \pi_k} \left[ \begin{array}{c} A_{\pi,C}(s,a) - \\ \frac{2\gamma \max[\mathbb{E}_{a\sim \pi_k}[A_{\pi,C}(s,a)]}{1-\gamma} \sqrt{\frac{1}{2} D_{KL}(\pi_k \| \pi)[s]} \end{array} \right] \quad (39)$$

Based on the bounds guaranteed, the following expected advantage function maximization problem can update the policy parameters iteratively:

$$\pi_{k+1} = \arg\max_\pi \mathbb{E}_{s\sim p, a\sim \pi_k}\left[ A_{\pi_k}(s,a) \right]$$
$$\text{s.t. } J_C(\pi_k) + \frac{1}{1-\gamma}\mathbb{E}_{s\sim p, a\sim \pi_k}\left[ A_{\pi_k,C}(s,a) \right] \leq d \quad (40)$$
$$D_{KL}(\pi \| \pi_k) \leq \delta$$

The problem can be solved by linearizing around the local policy. We will give detailed reformulation in the Part B.

*B. Consensus Multi-Agent Constrained Policy Optimization*

In the cooperative multi-agent reinforcement learning with local estimated global state value function, we have local estimated advantage functions in the environment for agents as:

$$A^i_{\pi_\theta^i}(s, a^i) = Q(s, \boldsymbol{a}) - V^i_\phi(s) \quad (41)$$

According to the function, we could derive the policy gradient algorithm for multi-agent reinforcement learning. We consider the local policy as $\pi^i_{\theta^i}$ for agent $i$. Then the gradient of the $J(\pi)$ with respect to local policy parameter $\theta^i$ is:

$$\nabla_{\theta^i} J(\pi) = \mathbb{E}_{s\sim p, a\sim \pi_\theta^i}[\nabla_{\theta^i} \log \pi^i_\theta(s|a^i) \cdot A^i_{\pi_\theta^i}(s, a^i)] \quad (42)$$

From the expression, we can conclude that the optimal policy can be obtained locally based on the estimation of the local advantage function. However, the local update may lead to the inaccurate estimation for the functions and the gradient with respect to the parameters so that we come up with communications between the agents when centralized controller is unavailable.

According to the networked multi-agent MDP mentioned in Section II, we assume that each agent estimates the global advantage function locally by the parameters $\phi^i$ of neural networks and shares the local information with their neighbors based on the communication graph. Then the estimation for the advantage function can reach a consensus and the policy of each agent can be improved through the policy gradient accurately. In practice, we always approximate the advantage functions through Generalized Advantage Estimation (GAE) algorithm as

$$(\hat{A}^i_t)^{GAE(\gamma, \lambda_{GAE})} = \sum_{l=1}^{\infty} (\gamma \lambda_{GAE})^l (r^i_t + \gamma V^i_\phi(s_{t+l+1}) - V^i_\phi(s_{t+l})) \quad (43)$$

So that we only need to parameterize the local estimated global state value functions.

Therefore, an actor-critic for the networked agent is proposed including the policy update for actor and value function update for critic. For the update of the critic's parameters, temporal difference (TD) learning is applied as

$$\delta^i_t = r^i_t + \gamma V^i_\phi(s_{t+1}) - V^i_\phi(s_t) \quad (44)$$
$$\tilde{\phi}^i_k = \phi^i_k + \beta_{\phi,k} \cdot \delta^i_t \cdot \nabla_{\phi^i} V^i(s; \phi^i) \quad (45)$$

To ensure the safe operation of the distributed demand management, we combine the constrained policy optimization with networked multi-agent reinforcement learning. We have similar local estimated cost advantage functions as (41) and they can also be estimated by GAE:

$$A^i_{\pi_\theta^i,C}(s, a^i) = Q^i_C(s, \boldsymbol{a}) - V^i_{\phi,C}(s) \quad (46)$$

Recall from CPO, agent maximizes the surrogate return subject to the surrogate cost constraint within a limited KL difference. In practice, the KL constraint is intractable due to the difficulties in computation of KL-divergence at every state so that we relax it by expected KL-divergence approximated from stochastic sampling. Therefore, the multi-agent optimization problem base on (40) can be rewrite as

$$\theta^i_{k+1} = \arg\max_{\theta^i} \mathbb{E}_{s\sim p, a^i \sim \pi^i_\theta}[A^i_{\pi^i_\theta}(s, a^i)]$$
$$\text{s.t. } J^i_C(\pi^i_{\theta,k}) + \frac{1}{1-\gamma}\mathbb{E}_{s\sim p, a^i \sim \pi^i_{\theta,k}}\left[ A^i_{\pi^i_{\theta,k},C}(s, a^i) \right] \leq d^i_t \quad (47)$$
$$\overline{D}_{KL}(\pi \| \pi_k) \leq \delta$$

For agent $i_h$, the estimated gradient of objective and cost are calculated as follows:

$$\boldsymbol{g}^{i_h}_k = \sum_{t=0}^{T} \nabla_{\theta^{i_h}_k} \log \pi^{i_h}_{\theta^{i_h}_k}(a^{i_h}_t | s_t) A^{i_h}_{\pi^{i_h}_\theta}(s, a^{i_h}) \quad (48)$$

$$\boldsymbol{b}^{i_h}_k = \sum_{t=0}^{T} \nabla_{\theta^{i_h}_k} \log \pi^{i_h}_{\theta^{i_h}_k}(a^{i_h}_t | s_t) A^{i_h}_{\pi^{i_h}_\theta,C}(s, a^{i_h}) \quad (49)$$

Accordingly, we will give detailed linearized reformulation for problem (47),

$$\theta_{k+1}^{i_h} = \arg\max_{\theta^{i_h}} (\boldsymbol{g}_k^{i_h})^T (\theta^{i_h} - \theta_k^{i_h})$$

$$\text{s.t. } J_C^{i_h}(\pi_{\theta_k^i}^{i_h}) - d^{i_h} + (\boldsymbol{b}_k^{i_h})^T(\theta^{i_h} - \theta_k^{i_h}) \le 0 \quad (50)$$

$$\frac{1}{2}(\theta^{i_h} - \theta_k^{i_h})^T \boldsymbol{H}_k^{i_h}(\theta^{i_h} - \theta_k^{i_h}) \le \delta$$

where $H_k^{i_h}$ is the hessian of the average KL-divergence of agent $i_h$ in iteration $k$. The optimization problem can be solved by primal-dual algorithms, as shown in Appendix. Then we could derive the solution to the problem as

$$\theta_*^{i_h} = \theta_k^{i_h} + \frac{\mu}{\lambda_*^{i_h}}(\boldsymbol{H}_k^{i_h})^{-1}(\boldsymbol{g}_k^{i_h} - \boldsymbol{b}_k^{i_h} \nu_*^{i_h}) \quad (51)$$

where $\lambda_*^{i_h}$ and $\nu_*^{i_h}$ are the solutions to the dual problem of (50) and $\mu$ is the step size for the update. Backtracking line search is applied in the algorithm to find an appropriate updating step size. If the problem is infeasible, we could take a TRPO step (52) on the cost surrogate.

$$\theta_{k+1}^{i_h} = \theta_k^{i_h} - \sqrt{\frac{2\delta}{(\boldsymbol{b}_k^{i_h})^T(\boldsymbol{H}_k^{i_h})^{-1}\boldsymbol{b}_k^{i_h}}}(\boldsymbol{H}_k^{i_h})^{-1}\boldsymbol{b}_k^{i_h} \quad (52)$$

Finally, based on the weight matrix which shows the weight on the information transmitted from agent $j$ to agent $i$ in the defined networked graph, we conduct a consensus step in the training process shown below:

$$\phi_{k+1}^i = \sum_{j \in \mathcal{N}} c(i,j) \cdot \tilde{\phi}_k^j \quad (53)$$

Note that the algorithm we propose is an on-policy reinforcement learning algorithm so that the trajectories are collected before the implementation. Therefore, the difficulties mentioned in the [27] may not affect the training.

The complete pseudocode of proposed algorithm is presented in **Algorithm 1**:

---
**Algorithm 1** CMACPO

1: **Input** distribution network topology, distributed generators' parameters, flexible loads' parameters, communication graph $\mathcal{G}$, historical data for loads and renewable energy resources, number of agents $n$, episodes $K$, hours per episode $T$.
2: **Initialize** actor network parameters $\theta_0$, local estimated global state value network parameters $\phi_0$, local cost value network parameters $\phi_{C,0}$, replay buffer $\mathcal{B}$, step size $\mu$.
3: **For** $k=0, 1, …, K-1$ **do**
4:    Sample a set of one-day data from historical data.
5:    **For** $t=0, 1, …, T-1$ **do**
6:      Observe the state $s_t$ from the environment.
7:      All load agents take actions $a_t$ based on $\pi$.
8:      Distribution network operator solve the problem (8)-(17), derive the price for each agent, calculate the nodal carbon intensity through (26) and then broadcast them to load agents to calculate $R_t$ and $C_t$.
9:      Store trajectories $(s_t, a_t, R_t, C_t, s_{t+1})$ into replay buffer $\mathcal{B}$.
10:     Estimate advantage function $A_{\pi_\theta}^i$ and cost advantage function $A_{\pi_\theta,C}^i$ based on corresponding value network with (43).
11:    **For** agent $i_h = i_1, …, i_n$ **do**
12:      Sample trajectories from $\mathcal{B}$, calculate the gradient of maximization objective $\boldsymbol{g}_k^{i_h}$ and the gradient of constraints $\boldsymbol{b}_k^{i_h}$ according to (48), (49) and the hessian of the average KL-divergence $\boldsymbol{H}_k^{i_h}$.
13:      Solve the problem (50).
14:      **If** feasible **do**
15:        obtain the optimal update (51)
16:      **else**
17:        apply a TRPO recovery (52)
18:      **end for**
19:    **For** agent $i_h = i_1, …, i_n$ **do**
20:      Update the local estimated state value function and local cost value function according to (44), (45)
21:    **end for**
22:    Do the consensus update for the state value function according to (53)
23: **end for**
24: Output optimal actor network parameters $\theta$, local estimated global state value network parameters $\phi$, local cost value network parameters $\phi_C$ and daily optimal flexible loads operation plan.
---

## IV. CASE STUDIES

The performance of proposed bi-level operation framework for demand management and algorithms is tested on the 33-bus and 123-bus distribution networks. Cases studies have been carried on a computer with an 8-core 2.20 GHz Intel i9-13900HX processor and 16GB of RAM. The algorithm is implemented based on Pytorch in Python 3.10 and Gurobi 10.0.

As shown in Fig. 2, the 33-bus distribution network is equipped with diesel generators located at bus 1 and 2, gas power plants located at bus 18 and 28 and renewable energy generation including wind turbines located at bus 3 and photovoltaic generators located at bus 7. Note that bus 0 is the common point between distribution network and external grid. The aggregated flexible loads for the operation are located at bus 12, 17, 19, 22, 25, regarded as five agents in the framework with the communication graph illustrated by red dashed lines shown in Fig. 2. The parameters for the dispatchable units are shown in Tab I. The daily data for the load consumption and the output of PV generators are sampled from the training data of [28] randomly. The voltage limit for the operation is [0.93, 1.05]. The daily carbon emission allowance for each agent is set based on the day-ahead allocation and carbon transactions and the total carbon emission allowance is 4.25 ton. The scheduling period $T$ is 24 hours and one step size $t$ is 1 hour.

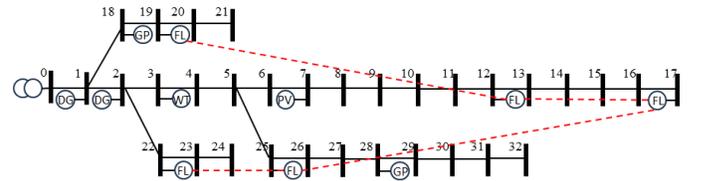

Fig. 2 Modified 33-bus distribution network

TABLE I. ALSAC ALGORITHM HYPERPARAMETERS

| Parameter | Value | Parameter | Value |
|---|---|---|---|
| $a_{DG,n}$ | 25.5/MW$^2$ | $P_{DG,n,\min}$ | 0MW |
| $b_{DG,n}$ | 205/MW | $P_{DG,n,\max}$ | 3MW |
| $c_{DG,n}$ | 50 | $P_{g,n,\min}$ | 0MW |
| $a_{g,n}$ | 10.5/MW$^2$ | $P_{g,n,\max}$ | 1.5MW |
| $b_{g,n}$ | 90/MW | $P_{grid,n,t,\max}$ | 5MW |
| $c_{g,n}$ | 12 | $\psi_g$ | 0.5 |
| $k_{re,n}$ | 55/MW | $\psi_{DG}$ | 0.9 |
| $\Delta$ /h | [4, 3, 5, 3, 4] | $\psi_{re}$ | 0.1 |

For the proposed CMACPO algorithm, the hyperparameters are shown in Tab II. Note that we conduct some effective tricks in our studies including Pop-Art[29], hybrid action space for policy network and Huber loss for critic networks.

TABLE II. ALSAC ALGORITHM HYPERPARAMETERS

| Parameter | Value |
| --- | --- |
| Size of hidden layer | [128, 32] |
| Activation function of hidden layer | ReLU |
| Initialization method | Orthogonal |
| Discount factor $\gamma$ | 0.95 |
| GAE factor $\lambda_{GAE}$ | 0.95 |
| Step size $\mu$ | 0.1 |
| Learning rate for $\phi$ | 5e-4 |
| KL-divergence threshold | 0.2 |

### A. Convergence Analysis

To illustrate the effectiveness of the proposed algorithm, we show some convergence results and compare it with other RL benchmark algorithms in this part.

When lacking of a prior knowledge and historical operation data in our framework, we consider about the comparisons between on-policy reinforcement learning including proximal policy optimization (PPO) developed from [26], constrained policy optimization (CPO) and multi-agent constrained policy optimization (MACPO). In PPO, when the carbon emission exceeds the limits, a penalty for the violation is added to the reward. In CPO, we assume that there is a central controller for the aggregated flexible loads and policy network, value network and cost value network are trained globally. In MACPO, there is no consensus step for the local estimated global state value function. Fig. 3 shows the reward convergence of the four algorithms. We could find that the proposed algorithm CMACPO has a great performance in the framework and the convergence curve is close to CPO, which is better than MACPO when the global state value function cannot be estimated precisely and PPO when there is no special treatment for the costs. Fig. 4 is the cost convergence curve showing that the cost in CMACPO merely exceeds the limits during the training process so that the low-carbon and safe load operation can be guaranteed. Note that the value in the curve is smoothed. Tab. III illustrates that the detailed convergence results of algorithms with random samples from test data based on the policy network for the aggregated flexible loads. In the proposed algorithm, the total profits for aggregated flexible loads are $839.3 and the reward converged after about 10000 episodes. There is no constraints violation rate, which is defined as the ratio of exceeding limit value to the carbon emission allowance and the total carbon emission is 4.20 ton.

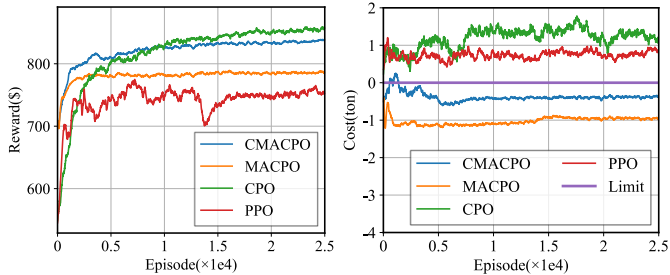

Fig. 3. Reward convergence    Fig. 4. Cost convergence

TABLE III. Detailed Comparison Results for Algorithms

| Method | Reward | Violation Rate | Carbon Emission |
| --- | --- | --- | --- |
| CMACPO | $ 839.3 | 0.0% | 4.20 ton |
| MACPO | $ 804.1 | 0.0% | 4.07 ton |
| CPO | $ 870.9 | 24.9% | 5.21 ton |
| PPO | $ 752.5 | 21.1% | 4.40 ton |

Fig. 5 represents the consensus convergence of local estimated global state value and the numerical number denotes the difference between agents' value estimation for the same state at a specific episode. We could find that the localized value function for each agent can reach consensus through the consensus step during the training process which could benefit the reward improvement, as concluded in Fig. 3.

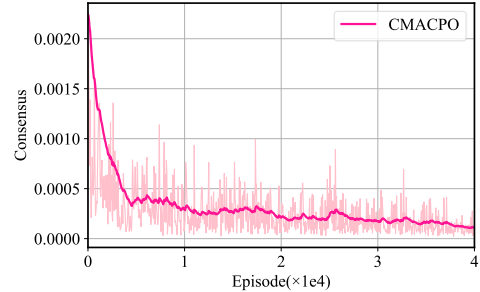

Fig. 5. Consensus convergence for CMACPO

### B. Result Analysis for the Operation Framework

In this part, we will show optimal scheduling plan for the hourly distributed low-carbon demand management in modified 33-bus distribution network with a specific sampled data. Fig. 6 denotes the environmental settings for the carbon intensity and wholesale price of the energy from external grid in a whole scheduling cycle and they are given at the beginning of each operation time step. We regard the distribution network operator as a price-taker in the wholesale market and the power exchange between the distribution network and main grid has little influence on the carbon intensity so that the carbon intensity and wholesale price is a constant for each time step. Fig. 7 shows the power output of wind turbines and PV in the environmental settings. Note that the two types of renewable energy resources present complementary characteristics in the time horizon. Both of them have the lower marginal cost and carbon intensity so that they can contribute to the reduction of carbon emission and payments for the distribution operator.

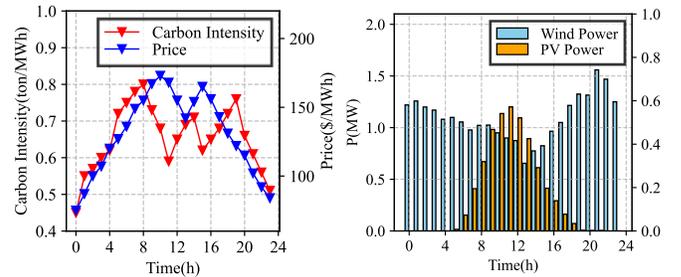

Fig. 6. Price and carbon intensity    Fig. 7. RES power output

Fig. 8 and 9 represents the hourly distributed flexible demand management results in the distribution network. We show the detailed load consumption and distribution locational marginal price for the specific bus in Fig. 8. In the proposed framework, the loss of the powerlines cannot be omitted so that the distribution locational marginal price is different between each bus. For the bus located at same branch, the distribution locational price is related, such as bus 12 and 17. Bus 25 is similar with 12 and 17 so that we don't show it here. The price for bus 22 is highly related to the wholesale market price and the price for bus 19 is highly related to the marginal cost for gas power plant which is located close to it.

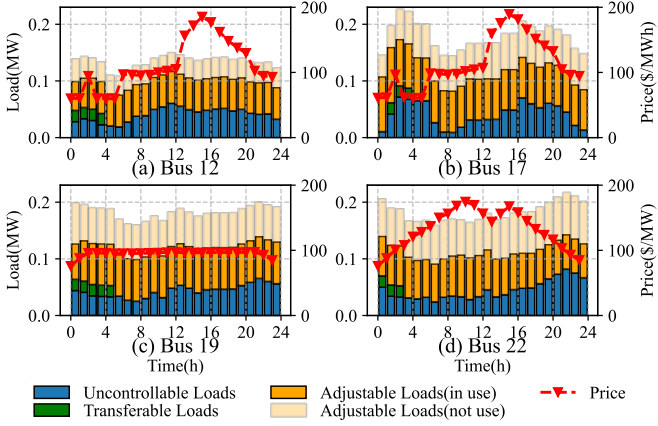

Fig. 8. Flexible load profile and DLMP

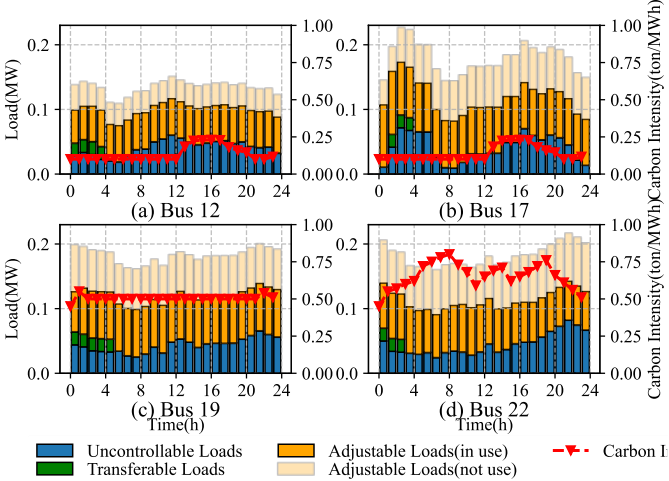

Fig. 9. Flexible load profile and carbon intensity

Similarly, we can make conclusion for the nodal carbon intensity with flexible load located in Fig. 9. For bus 22, the carbon intensity is corresponding with main grid and for bus 19, the carbon intensity is highly related with gas power plant. For bus 12 and 17, the carbon intensity is much lower than the other two. According to the definition of carbon intensity, it implies a 'weighted sum' of carbon intensity different types of generators so that the carbon intensity is low when the power consists of generation from renewable energy resources. For the flexible loads, we could find that when the price and the carbon intensity is relatively low, the ratio of adjustable load in use is higher. Also, the transferable loads are in use during the period to reduce the total payments. The phenomena happen in the carbon intensity similarly.

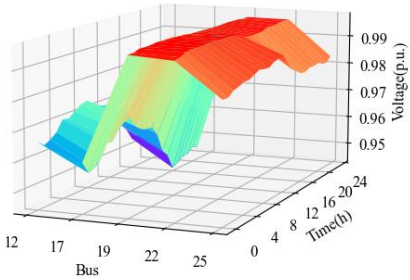

Fig. 10. Voltage profile in modified 33-bus system

Security operation for the distribution network in the proposed operation framework is illustrated in the Fig. 10. We can find that the voltage for the flexible load bus is limited in the safe range set before which demonstrate that the proposed algorithm for the framework can efficiently handling the constraints in the operation of distribution network.

Then we investigate the efficiency of the carbon emission reduction in the proposed framework by comparing with other two cases. *Case I*: The loads are operated without flexibility and carbon emission allowance. *Case II*: The loads are operated with flexibility but without carbon emission allowance.

However, for Case I, the carbon emission exceeds the limits about 42.14% and the reward is the lowest due to the inflexibility. For Case II, the reward is higher because of the flexible operation without any restricts. While the optimal operation in the case may lead to more carbon emission due to the lack of carbon emission allowance. The proposed framework is a tradeoff between the operation reward and carbon emission reduction. The detailed operation results can be found in Tab. IV.

TABLE IV. Detailed Comparison Results for Cases

| Case | Reward | Violation Rate | Carbon emission |
|---|---|---|---|
| Proposed | $ 839.3 | 0.0% | 4.196 ton |
| Case I | $ 596.5 | 42.14% | 6.041 ton |
| Case II | $ 973.5 | 9.06% | 4.635 ton |

### C. Scalability

With the increasing of aggregated flexible loads, centralized algorithm is more difficult to conduct due to the huge computational burden and high requirement of communication. The proposed algorithm is scalable in the extended operation setting of distribution network. To illustrate the scalability of the multi-agent reinforcement learning algorithms comparing with the single-agent centralized reinforcement learning algorithm, we implement our CMACPO, MACPO and CPO on the modified 123-bus distribution network.

In the settings, we have 10 aggregated flexible loads located at bus 11, 24, 33, 39, 51, 66, 75, 83, 96 and 113. The modified 123-bus distribution network is equipped with diesel generators located at bus 2, 8, 18 and 86, gas power plants at bus 13, 47 and 101 and renewable energy generation including wind turbines at bus 25 and photovoltaic generators at bus 57, 76, 89 and 108. The total carbon emission allowance is set as 12.8 ton. Other settings are similar to the modified 33-bus distribution network test case in part A. We show the convergence result of the three algorithms tested on the environment settings. From Fig. 11 and 12, we could find that under the similar algorithm settings, the proposed algorithm has better convergence result and less convergence iterations than the other algorithms. As concluded in the Tab. V, the training reward of the proposed algorithm is $1542.0, which is the highest reward when the carbon constraints are satisfied.

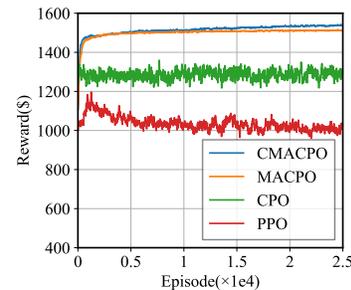 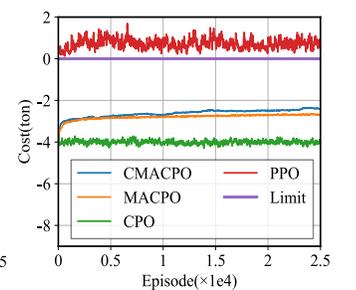

Fig. 11. Reward convergence    Fig. 12. Cost convergence

TABLE V. Detailed Comparison Results for Algorithms

| Method | Reward | Violation rate | Carbon Emission |
|---|---|---|---|
| CMACPO | $ 1542.0 | 0.0% | 10.47 ton |
| MACPO | $ 1509.3 | 0.0% | 10.21 ton |
| CPO | $ 1419.0 | 0.0% | 9.22 ton |
| PPO | $ 1031.1 | 0.86% | 12.85 ton |

Contrary to the proposed algorithms, centralized RL algorithm such as PPO and CPO cannot handle high dimensions appropriately so that the reward curve shows instability and few improvements in the training process.

Therefore, the proposed algorithm can be applied in the large-scale distribution system operation framework with more aggregated flexible load agents under the limits of carbon emission allowance, and the convergence result and optimal dispatching plan validates the efficiency and scalability of the proposed algorithm.

## V. CONCLUSION

In this paper, a bi-level distributed low-carbon demand management framework in the distribution network is proposed that aggregated loads send their demand and receive distribution local marginal price and carbon intensity from cost-minimization distribution system operator in the lower level to maximize their profits. In the framework, the demand-side carbon emission allowance is considered to transfer the responsibility of carbon emission from generation to demand to realize the reduction of carbon emission. Due to the incomplete information exchange and hourly distributed operated requirements with uncertain renewable energy resources and uncontrollable loads, we formulate the problem into a networked multi-agent constrained Markov Decision Process and introduce a networked multi-agent constrained policy optimization to solve it. The efficiency and effectiveness of proposed framework and algorithm on profit maximization with the carbon emission limit and safe operation guarantee is validated based on the modified 33-bus distribution network and 123-bus distribution network, compared with existing reinforcement learning algorithm and other operation cases.